\documentclass[aip,reprint]{revtex4-1}

\usepackage{graphicx}% Include figure files
\usepackage{dcolumn}% Align table columns on decimal point
\usepackage{bm}% bold math
\usepackage{upgreek}
\usepackage[utf8]{inputenc}
\usepackage[T1]{fontenc}
\usepackage{mathptmx}
\usepackage{amsmath}

\draft % marks overfull lines with a black rule on the right

\begin{document}

\preprint{AIP/123-QED}

\title[A silicon source of heralded single photons at 2 $\mu$m. ]{A silicon source of heralded single photons at 2 $\mu$m.}
% Force line breaks with \\

\author{S. Signorini}
 \email{stefano.signorini-1@unitn.it} \affiliation{Nanoscience Laboratory, Department of Physics, University of Trento, Via Sommarive 14, 38123, Trento, Italy}%Lines break automatically or can be forced with \\

\author{M. Sanna}%
\affiliation{Nanoscience Laboratory, Department of Physics, University of Trento, Via Sommarive 14, 38123, Trento, Italy}
 
\author{S. Piccione}%
\affiliation{Nanoscience Laboratory, Department of Physics, University of Trento, Via Sommarive 14, 38123, Trento, Italy}

%\author{M. Bernard}%
%\affiliation{Centre for Materials and Microsystems, Fondazione Bruno Kessler, 38123, %Trento, Italy}

\author{M. Ghulinyan}%
\affiliation{Centre for Sensors and Devices, Fondazione Bruno Kessler, 38123, Trento, Italy}

\author{P. Tidemand-Lichtenberg}%
\affiliation{Department of Photonics Engineering, DTU Fotonik, Technical University of Denmark, Roskilde, 4000, Denmark}

\author{C. Pedersen}%
\affiliation{Department of Photonics Engineering, DTU Fotonik, Technical University of Denmark, Roskilde, 4000, Denmark}

\author{L. Pavesi}%
\affiliation{Nanoscience Laboratory, Department of Physics, University of Trento, Via Sommarive 14, 38123, Trento, Italy}

%\affiliation{ 
%Authors' institution and/or address%\\This line break forced with \textbackslash\textbackslash
% Use the \preprint command to place your local institutional report number 
% on the title page in preprint mode.
% Multiple \preprint commands are allowed.
%\preprint{}

%\title{} %Title of paper

% repeat the \author .. \affiliation  etc. as needed
% \email, \thanks, \homepage, \altaffiliation all apply to the current author.
% Explanatory text should go in the []'s, 
% actual e-mail address or url should go in the {}'s for \email and \homepage.
% Please use the appropriate macro for the type of information

% \affiliation command applies to all authors since the last \affiliation command. 
% The \affiliation command should follow the other information.

%\author{}
%\email[]{Your e-mail address}
%\homepage[]{Your web page}
%\thanks{}
%\altaffiliation{}
%\affiliation{}

% Collaboration name, if desired (requires use of superscriptaddress option in \documentclass). 
% \noaffiliation is required (may also be used with the \author command).
%\collaboration{}
%\noaffiliation

%\date{\today}

\begin{abstract}
Mid infrared integrated quantum photonics is a promising platform for applications in sensing and metrology. However, there are only few examples of on-chip single photon sources at these wavelengths. These have limited performances with respect to their C-band counterparts. In this work, we demonstrate a new approach to generate heralded single photons in the mid infrared on a silicon chip. By using a standard C-band pump, the inter-modal spontaneous four wave mixing enables the generation of the herald idler at 1259.7 nm and the heralded signal at 2015 nm. The idler photon is easily detected with a common infrared single photon detector while the signal photon is upconverted to the visible before its detection. In this way, we are able to operate a mid infrared source without the need of mid infrared detectors and laser sources. By measuring a heralded $g^{(2)}$ of $0.23 \, \pm \, 0.08$ we demonstrate the single photon behaviour of the source as well as the feasibility of multi-photon coincidence measurements beyond 2 $\mu$m with our setup. The source exhibits a high intrinsic heralding efficiency of $(59 \, \pm \,5)\%$, a maximum coincidence to accidental ratio of $40.4 \, \pm \, 0.9$ and a generation probability of $\left( 0.72 \, \pm \, 0.10 \right)$ W$^{-2}$.
%, to date the highest performance for a mid infrared source in silicon.
\end{abstract}

% insert suggested PACS numbers in braces on next line

\maketitle %\maketitle must follow title, authors, abstract and \pacs

% Body of paper goes here. Use proper sectioning commands. 
% References should be done using the \cite, \ref, and \label commands
\section{Introduction} \label{sec:introduction}
Mid infrared (MIR) light (2 - 15 $\upmu$m) is of importance in a wide range of technological applications. Free space telecommunication \cite{su201810}, LIDAR \cite{weibring2003versatile}, environmental monitoring \cite{fix2016upconversion}, medicine and biology \cite{evans2007chemically,bellisola2012infrared,potter2001imaging,miller2013ftir} are only few of the several fields where MIR optics plays a role. In particular, gas sensing exploits the strong absorption bands in the MIR \cite{popa2019towards} to enhance remarkably the sensitivity of absorption spectroscopy measurements \cite{petersen2014mid,vainio2016mid,ghorbani2017real}. Despite the great interest in developing MIR applications, these are still hindered by immature optical MIR devices. Quantum optics offers new solutions to mitigate such limitations. Sub-poissonian light can be used to beat the shot noise limit \cite{brida2010experimental,whittaker2017absorption}. Entangled photons have been used to demonstrate new imaging and spectroscopy techniques able to get rid of detection technology limitations, namely ghost imaging \cite{pittman1995optical,morris2015imaging} or undetected photon measurement \cite{lemos2014quantum,kalashnikov2016infrared,vergyris2020two}. To enable quantum enhanced MIR metrology leveraging these quantum based measurement strategies, a source of single or entangled photons beyond 2 $\upmu$m is required. Up to now, these techniques have been investigated only with bulky, alignment tolerant and expensive instrumentation, based on free space nonlinear crystals \cite{kalashnikov2016infrared,prabhakar2020two}. To develop feasible, robust and affordable quantum technologies, miniaturization and cost effectiveness are crucial. Such requirements can be met by means of integrated photonics. In particular, silicon photonics integrated circuits are characterized by mature CMOS (complementary metal oxide semiconductor) fabrication technology, which allows for robust, stable, low power consuming and efficient light manipulation at the chip scale \cite{lockwood2010silicon}. On-chip MIR quantum measurements would enable efficient and cost effective sensors, boosting the development of MIR and quantum technologies.
 Recently, an on-chip silicon-on-insulator (SOI) source of MIR pairs has been reported \cite{rosenfeld2020mid}. However, in this work a pump in the MIR is used, and both the paired photons are beyond 2 $\upmu$m, thus requiring specific MIR technologies for both the pump and the detection. Recently, we demonstrated that inter-modal spontaneous four wave mixing (SFWM) can be used in silicon waveguides to generate correlated pairs with one photon in the near infrared (NIR) and the other in the MIR by using a standard C-band pump \cite{signorini2018intermodal,signorini2019silicon}. However, we never detected the MIR correlated photon. Instead we inferred its existence by measuring the high energy photon in the pair.\\
In this work, we demonstrate a SOI waveguide source of heralded MIR single photons based on inter-modal SFWM, peforming the MIR detection by means of an upconversion system \cite{mancinelli2017mid}. The herald photon lays in the NIR, where it can be efficiently detected with traditional InGaAs single photon avalanche photodiodes (SPADs). Moreover, the photons are generated in discrete bands, thus removing the need for narrow band filters to select the operating wavelengths of signal and idler. As a result, the heralding efficiency is increased with respect to traditional intra-modal SFWM, as witnessed by the measured intrinsic heralding efficiency $\eta_I = 59(5) \, \%$. The large detuning of the generated photons is also beneficial for the pump and Raman noise rejection, that can be easily removed with broadband filters. The pump is a standard 1550 nm pulsed laser. Therefore, we do not require MIR technologies to operate a source beyond 2 $\mu$m. We assessed the single photon behaviour of the source by measuring a heralded $g^{(2)}_h(0)$ of 0.23(8). We monitored the idler-signal coincidences, reporting a maximum coincidence to accidental ratio of 40.4(9), exceeding the performance of current integrated sources of MIR heralded photons \cite{rosenfeld2020mid}.\\ The paper is organized in the following way:
in section \ref{sec:setup} we describe the chip design and the experimental setup. In section \ref{sec:data_anal} our approach to data analysis is extensively described. In section \ref{sec:results} the results relative to the source characterization are reported. Section \ref{sec:conclusions} concludes the paper.

\begin{figure*}[t]
\includegraphics[width=\textwidth]{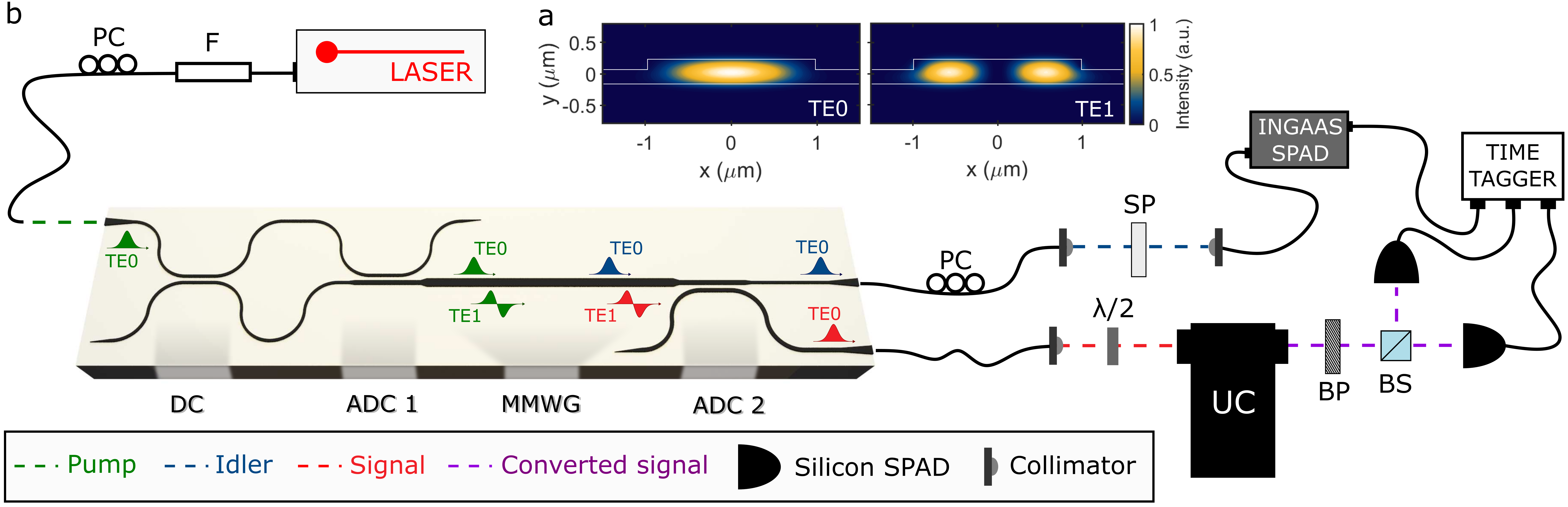}
\caption{\label{fig:setup} a) Simulated intensity profiles of the TE0 and TE1 spatial modes in the multimode waveguide. b) Experimental setup used in the experiment. For the pump (green) we used a pulsed laser at 1550.3 nm (40 ps pulse width, 80 MHz repetition rate), which, after passing through a band pass filter (F) and a polarization controller (PC), is coupled to the chip via a tapered lensed fiber. The chip schematics is shown in the bottom part. On the chip, after a 3-dB directional coupler (DC), half of the pump remains on the TE0, while the other half is converted to the TE1 via an asymmetric directional coupler (ADC1) (92$\%$ efficiency). In this way, the pump reaches the multimode waveguide (MMWG) equally splitted on the TE0 and TE1 modes. In the MMWG, the inter-modal SFWM process generates the idler (blue) and signal (red) photons in the TE0 and TE1 modes respectively. The signal is then converted to the TE0 via another asymmetric directional coupler (ADC2). In this way, idler and signal can be easily separated on chip. Idler and signal are then out-coupled from the chip via two tapered lensed fibers. Pump residual and Raman noise are rejected from the idler beam by means of a short pass filter (SP) with a cut-off wavelength of 1335 nm. The idler is then detected via an InGaAs SPAD (ID Quantique IDQ210), triggered by the pump, with a gate width of 1.90 ns. The signal, after being out-coupled from the chip, is polarization rotated through a free space half-wave plate $\left( \lambda / 2 \right)$ and upconverted to the visible through an upconverter system (UC). The UC includes a long pass filter with a cut-on wavelength of 1900 nm, which rejects the C-band pump. To be noticed that the UC introduces noise photons collinear to the upconverted signal and centered at the same wavelength. A bandpass filter (BP) is used to filter away part of this noise, without filtering the upconverted signal (purple). Then, the signal photons are analyzed by means of a Hanbury Brown and Twiss (HBT) interferometer. The HBT interferometer is composed by a 50/50 beam splitter (BS) with two visible silicon SPADs (Excelitas SPCM-AQRH-12) monitoring the BS reflection and transmission ports. The visible SPADs are used in free-running mode. A time tagging unit (Swabian Time Tagger 20) is used to monitor individual singles and coincidences between the three detectors.
}
\end{figure*}

\section{Chip design and experimental setup}\label{sec:setup}
Conventional intra-modal SFWM involves only one waveguide mode in the conversion of two input pump photons into an idler photon and a signal photon. On the contrary, inter-modal SFWM leverages the different chromatic dispersions of different optical spatial modes of a photonic waveguide to achieve phase matching \cite{signorini2018intermodal}. Different modal combinations are possible, depending on the waveguide cross-section, which also determines the generated signal and idler wavelengths. In this work, we use the transverse electric (TE) fundamental (TE0) and first (TE1) waveguide modes in a rib SOI waveguide. The waveguide has a width of 1.95 $\upmu$m and a height of 0.190 $\upmu$m over a 0.3 $\upmu$m thick slab. The waveguide length is 1.5 cm. The waveguide and the slab are in silicon, while the top and bottom claddings are in silica. The simulated intensity profiles of the TE0 and TE1 modes are shown in Fig. \ref{fig:setup}a.

\begin{figure}[!t]
\includegraphics[width=\columnwidth]{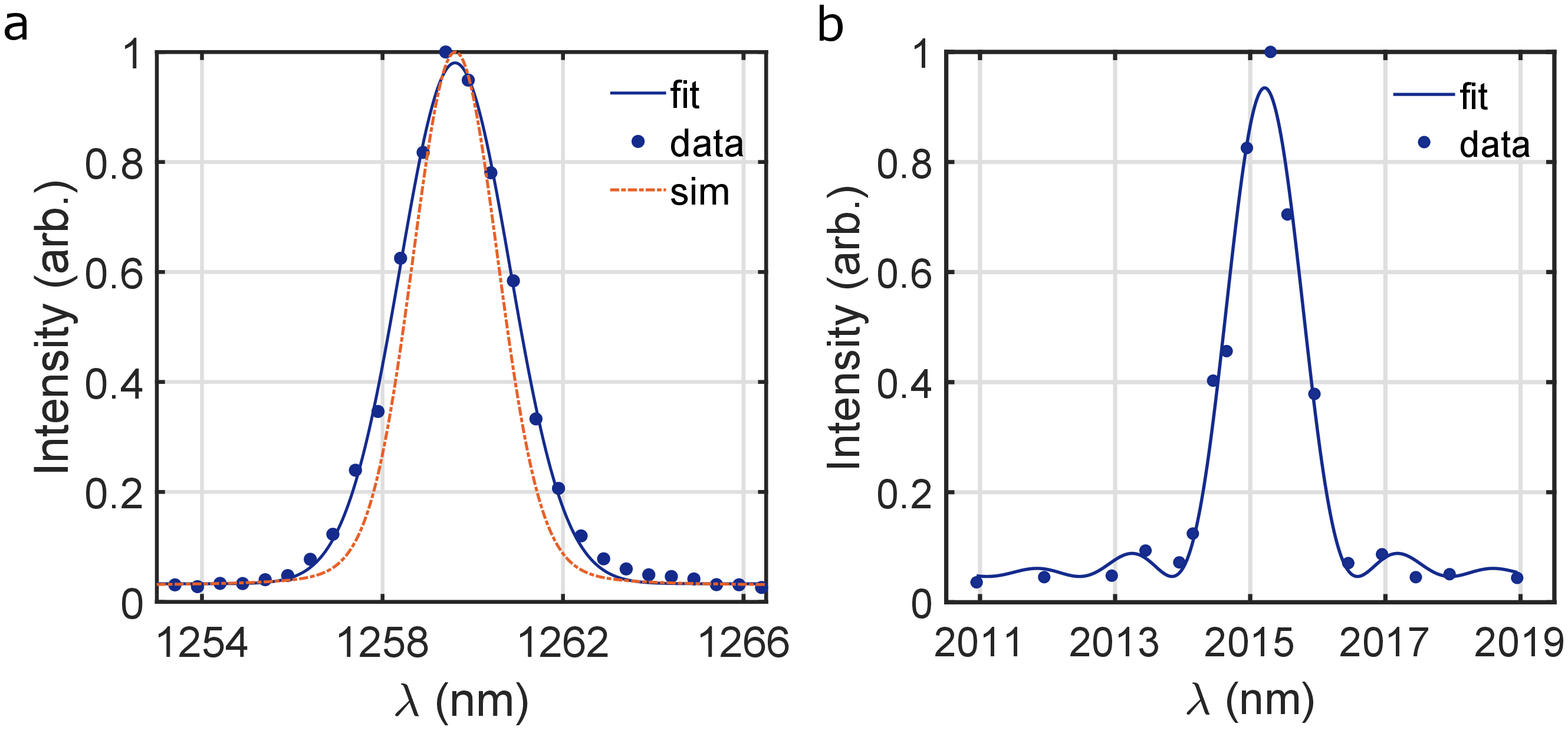}% Here is how to import EPS art
\caption{\label{fig:profiles} a) Measured intensity spectrum of the idler beam. The fit has been made with a gaussian function, showing a FWHM of $2.87 \, \pm \, 0.07$ nm. This measurement is affected by the transfer function of the monochromator used to perform the measurement, which enlarges the actual bandwidth of the generation. We simulated the idler spectrum considering also the widening due to the monochromator (orange dashed line). To evaluate the actual bandwidth of the idler (2.0 $\pm$ 0.3 nm) we deconvolved the response function of the monochromator. b) Measured spectral response of the upconverter. The response has been fitted by a squared sinc function, as expected for a sum frequency generation process. The FWHM is $1.15 \, \pm \, 0.12$ nm. }
\end{figure}

The inter-modal combination used in our work involves the pump on both the TE0 and TE1, the idler on the TE0 and the signal on the TE1. A peculiar advantage of intermodal SFWM is the generation of the signal and idler photons on different waveguide modes. In this way, idler and signal can be easily separated with high efficiency through an on-chip mode converter. 
The experimental setup is detailed in Fig. \ref{fig:setup}b.
The upconverter (UC) is constituted by a continuous wave (CW) laser cavity, where a  Nd:YVO$_4$ pumped intra-cavity periodically poled lithium niobate (PPLN) allows for sum-frequency generation (SFG) between the intra-cavity laser (1064 nm) and the input MIR photons. We used a PPLN from HC Photonics with a length of 25 mm, tuned in temperature to upconvert the MIR signal at 2015 nm to the visible at 696 nm. The UC used is the same of Mancinelli et al. \cite{mancinelli2017mid}, though tuned at the wavelengths of interest here. The transfer function of the UC is reported in Fig. \ref{fig:profiles}b, showing a full width at half maximum (FWHM) of $1.15 \, \pm \, 0.12$ nm.
We used a pump pulsed laser centered at $1550.30 \, \pm \, 0.05$ nm with 40 ps pulse width and 80 MHz repetition rate. The generated idler spectrum is reported in Fig. \ref{fig:profiles}. We measured a discrete band centered at 1259.7 $\pm$ 0.5 nm, with a FWHM of 2.0 $\pm$ 0.3 nm. The measured FWHM of the idler is compatible with the simulated one of 1.81 nm, as shown in Fig. \ref{fig:profiles}a. According to the energy conservation, the signal is generated at 2015.2 $\pm$ 1.5 nm. From the measured idler bandwidth we estimated a FWHM of 5.1 $\pm$ 0.8 nm for the signal. Therefore, the UC filters the signal photons according to the spectrum shown in Fig. \ref{fig:profiles}b.
 
\section{Data analysis}\label{sec:data_anal}
With SFWM, the detection probabilities per pulse for the idler ($p_i$), signal ($p_s$), coincidences ($p_{si}$) and accidentals ($p_{acc}$) are quadratic with the pump power $P$. In the limit of low transmission efficiencies for the signal and idler \cite{harada2009frequency}, they can be written as
\begin{subequations}\label{eq:probabilities}
\begin{equation} \label{prima_pi}
p_i = \xi P^2 \eta_i + d_i,
\end{equation}
\begin{equation}\label{prima_ps}
p_s = \xi P^2 \eta_s + d_s,
\end{equation}
\begin{equation}\label{prima_pcc}
p_{si} = \xi P^2 \eta_i \eta_s,
\end{equation}
\begin{equation}\label{prima_pacc}
p_{acc} = p_i p_s,
\end{equation}
\end{subequations}
where $\xi$ is the generation probability per pulse per squared unit power \cite{rosenfeld2020mid}, $\eta_i, \, \eta_s$ are the total transmission efficiencies for the idler and signal channels (from generation to detection), $d_i, \, d_s$ are the dark count probabilities per pulse for the idler and signal respectively. Eq. \eqref{prima_pcc} refers to net coincidences, thus without accidentals. In eqs. \eqref{eq:probabilities}, noise photons coming from the pump residual and Raman scattering, typically linear with the pump power, have not been considered, being negligible in our experimental setup. Singles and coincidence rates can be calculated by multiplying the probabilities in eqs. \eqref{eq:probabilities} by the repetition rate $R_p$ of the pump laser.
Together with SFWM other nonlinear phenomena take place in the waveguide. Two photon absorption (TPA), cross two photon absorption (XTPA) and free carrier absorption (FCA) have to be modelled properly in order to recover the actual generation and transmission efficiency of the pairs. TPA, XTPA and FCA play an important role in increasing the losses in the waveguide for both the pump and the generated photons. As a result, the detection probabilities are no longer quadratic with the input pump power \cite{boyd2019nonlinear}. A further effect is the nonlinearity of the idler detector. To model the linear and nonlinear losses affecting pump, signal and idler photons, we solved the differential equations for the pulse propagation involving TPA, FCA and propagation losses, assuming that the pump power is equally split on the TE0 and TE1 modes \cite{borghi2017nonlinear}. According to this modeling we can rewrite eqs. \eqref{eq:probabilities} as 
\begin{subequations} \label{p_average}
\begin{equation} \label{pcc_average}
    p_{si} \simeq \xi \Bar{P}_p^2 \Bar{\eta}_i \Bar{\eta}_s \eta_{ND} \equiv \bar{p}_{si},
\end{equation}
\begin{equation} \label{pi_average}
    p_i \simeq \left( \xi \Bar{P}_p^2 \Bar{\eta}_i + d_i \right) \eta_{ND} \equiv \bar{p}_{i},
\end{equation}
\begin{equation} \label{ps_average}
    p_s \simeq \xi \Bar{P}_p^2 \Bar{\eta}_s + d_s \equiv \bar{p}_{s},
\end{equation}
\begin{equation}\label{pacc_average}
p_{acc} \simeq \bar{p}_i \bar{p}_s \equiv \bar{p}_{acc},
\end{equation}
\end{subequations}
where 
\begin{subequations}
\begin{equation}
    \Bar{P}_p = \sqrt{\frac{1}{L} \int_0^L P_p^2(z) dz},
\end{equation}
\begin{equation}
    \bar{\eta}_j = \bar{\eta}_j^{on} \eta_j^{off},
\end{equation}
\begin{equation}\label{etajbar}
    \Bar{\eta}^{on}_j = \frac{1}{L} \int_0^L \eta^{on}_j(z) dz,
\end{equation}
\end{subequations}

where $j=i,s$, $L$ is the waveguide length, $P_p(z)$ is the on-chip pump power along the waveguide, $\eta^{on}_j(z)$ is the transmission efficiency for a photon generated at $z$ along the waveguide accounting only for the linear and nonlinear on-chip losses, $\eta^{off}_j$ is the transmission efficiency accounting only for the losses occurring off chip (fiber-chip coupling, filtering) and $\eta_{ND}$ models the nonlinear response of the idler detector. 
Details about the derivation of eqs. \eqref{p_average} are reported in Supplementary material.

\section{Results} \label{sec:results}
\subsection{Generation probability and heralding efficiency}\label{brightness}
To monitor the coincidences between signal and idler, we used a start-and-stop detection system, using the idler as the start trigger and the signal as the stop detection \cite{signorini2020chip}. Coincidences are evaluated within a coincidence window $\Delta t_c$. To be noticed that while for the idler channel the detection rates (both signal and dark counts) are fixed by the detection gate width of the idler detector (1.90 ns), for the signal the rates depend on the coincidence window used in post processing. Therefore, given $R_{dc,i} = 620 \, \textrm{cps}$ and $R_{dc,s} = 2150 \, \textrm{cps}$ the dark count rates at the idler and signal detectors, 
\begin{equation}
d_i = R_{dc,i}/R_p = 7.75\times 10^{-6},
\end{equation}
while
\begin{equation}
d_s = 1 - \textrm{e}^{-R_{dc,s} \Delta t_c},
\end{equation}
considering a Poisson distribution for the signal noise (SPAD dark counts and UC noise). 

In order to fit the measured rates and retrieve the generation probability $\xi$, we can reduce eqs. \eqref{p_average} to
\begin{subequations}\label{y}
\begin{equation}
    y_i = \frac{\bar{p}_i -  \eta_{ND} \, d_i}{\bar{\eta}_i^{on} \eta_{ND} } = \xi \bar{P}_p^2 \eta_i^{off} = a_i \bar{P}_p^2, 
\end{equation}
\begin{equation}
    y_s = \frac{\bar{p}_s - d_s}{ \bar{\eta}_s^{on} } = \xi \bar{P}_p^2 \eta_s^{off} = a_s \bar{P}_p^2,
\end{equation}
\begin{equation}
    y_{si} = \frac{\bar{p}_{si}}{\bar{\eta}_i^{on} \eta_{ND} \bar{\eta}_s^{on}} = \xi \bar{P}_p^2 \eta_i^{off} \eta_s^{off} = a_{si} \bar{P}_p^2, 
\end{equation}
\end{subequations}
with $a_i = \xi \eta_i^{off}$, $a_s = \xi \eta_s^{off}$, $a_{si} = \xi \eta_i^{off} \eta_s^{off}$. $y_i$, $y_s$, $y_{si}$ can be calculated from the measured singles, coincidence and noise rates and from the simulated $\bar{\eta}^{on}_j$ and the measured $\eta_{ND}$ (see Supplementary material). Modeling exactly the nonlinear losses is a non trivial task, being the nonlinear parameters highly variable with the fabrication process and the geometry used. Therefore, we fit $y_i$, $y_s$, $y_{si}$ for an input power $<$ 0.5 W (i.e. $\bar{P}_p < 0.4$ W), where the nonlinear losses are not the dominant ones. We use $f(x) = a x^2 + b$ as the fitting function, retrieving $a_i$, $a_s$ and $a_{si}$. In this way, we can evaluate $\xi$ (in units of $W^{-2}$ of peak power) and the off-chip transmissions, resulting in
\begin{subequations} \label{fit_results}
\begin{equation}
    \xi = \frac{a_i \, a_s}{a_{si}} = \left( 0.72 \pm 0.10 \right) W^{-2}, 
\end{equation}
\begin{equation}
    \eta_i^{off} = \frac{a_{si}}{a_s} = \left( 2.81 \pm 0.17 \right)\times 10^{-3}, 
\end{equation}
\begin{equation}
    \eta_s^{off} = \frac{a_{si}}{a_i} = \left( 3.97 \pm 0.20 \right)\times 10^{-4},
\end{equation}
\end{subequations}
where we used $\Delta t_c = 1.1$ ns (3$\sigma$ bin width) and the uncertainties are evaluated at 1 standard deviation of the fitting coefficients. Details about the nonlinear parameters and propagation losses used in the model are reported in Supplementary materials. From these results we calculate the intrinsic heralding efficiency $\eta_I$ as \cite{signorini2020chip}
\begin{equation}
    \eta_I = \frac{R^{net}_{si}}{\left( R_i-R_{dc,i}\right) \, \bar{\eta}_s^{off}} = 59 \pm 5 \, \%,
\end{equation}
where $R^{net}_{si}$ is the measured net coincidence rate and $R_i$ is the measured idler rate. By normalizing for the signal channel losses, the $\eta_I$ allows to compare different sources only on the bases of their intrinsic properties, getting rid of the setup used. Our high value comes from the low on-chip signal losses and the moderate filtering losses to select the signal wavelength. The heralding efficiency can be further improved by optimizing the matching between the signal and UC bandwidths.

\subsection{Coincidence to accidental ratio}
\begin{figure}[h!]
\includegraphics[width=\columnwidth]{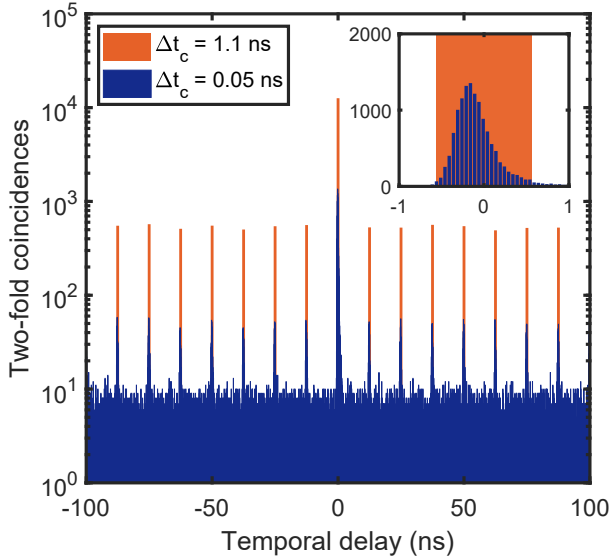}% Here is how to import EPS art
\caption{\label{fig:car_meas} 
Two-fold coincidences as a function of the delay $\delta t$ between idler (start) and signal (stop) detections. We collect the events with a coincidence window of 0.05 ns (blue). In post processing, we use a larger coincidence window, here 1.1 ns (orange), in order to take into account the majority of the coincidence events. The coincidence peak is the highest one, placed at $\delta t = 0$ ns. The laser repetition period is clearly visible from the accidental peaks. In the inset, we focus on the zero-delay bin, comparing the coincidence peak shape with the post processing coincidence window.
}
\end{figure}

\begin{figure}[h!]
\includegraphics[width=\columnwidth]{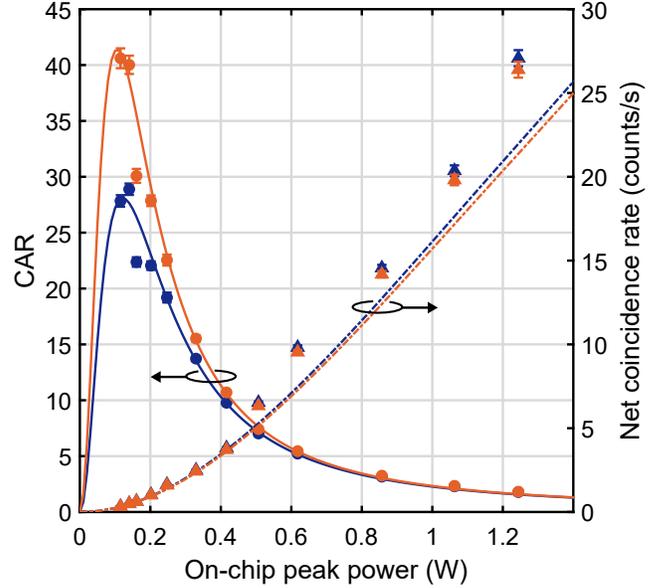}% Here is how to import EPS art
\caption{\label{fig:car} 
Measured CAR (circles) and net coincidence rates (triangles) with $\Delta t_c = 1.1$ ns (orange) and $\Delta t_c = 2$ ns (blue). The data are reported versus the on-chip peak pump power. The experimental points are compared with the simulated values for both the CAR (solid lines) and the net coincidence rates (dashed lines). With $\Delta t_c = 1.1$ ns the CAR is remarkably higher with respect to the 2 ns bin, with only a limited reduction in the coincidence rate. The better performance obtained with the smaller $\Delta t_c$ is due to the lower noise integrated within the coincidence bin.
}
\end{figure}

To quantify the efficiency of coincidence detection, the coincidence-to-accidental ratio (CAR) is used. CAR is analogous to a signal-to-noise ratio comparing the rate of true coincidences with the accidental ones. True coincidences come from simultaneous detection of a signal and an idler belonging to the same pair. Coincidences between signals and idlers belonging to different pairs or coincidences with noise photons or dark counts contribute to the accidentals \cite{signorini2020chip,harada2009frequency}. The measurement of CAR is carried out with the start-stop coincidence detection described in sec. \ref{brightness}. We used the setup in Fig. \ref{fig:setup}b with a single visible SPAD at the output of the UC after removing the beams splitter. In fact, the CAR does not involve the intra-beam correlations. As shown in Fig. \ref{fig:car_meas}, the coincidences occur with a temporal delay $\delta t$ = 0 ns. The other peaks, spaced with the laser repetition period, are due to accidentals. Please notice that the zero-delay peak includes also  accidental coincidences. 
Therefore, the CAR is evaluated as
\begin{equation}
\textrm{CAR} = \frac{\textrm{coincidence counts}}{\textrm{accidental counts}} = \frac{N^{raw}_{si} - N_{acc}}{N_{acc}}, 
\end{equation}

with $N^{raw}_{si}$ the total coincidence counts falling in the zero delay bin and $N_{acc}$ the accidental counts, evaluated as the average over all the accidental peaks. The true coincidences, also called as net coincidences, are calculated as $N_{si}^{net} = N_{si}^{raw}-N_{acc}$. Depending on the $\Delta t_c$ used, the ratio between coincidence and accidentals in the individual bin changes, changing the CAR. In Fig. \ref{fig:car} we report the measured CAR and the corresponding net coincidences as a function of the on-chip peak pump power. To be noticed that the peak power in the plot is the power at the input of the multimode waveguide after fiber-chip coupling losses, it is not $\bar{P}_p$. We report the results with a coincidence window of 1.1 ns and of 2 ns. With the 1.1 ns window the CAR is higher, with a maximum of 40.4(9) at 115 mW. At this power the rate of net coincidences is 0.316(3) cps. The net coincidences are almost the same for the two windows, demonstrating that with the larger coincidence window we are mainly introducing noise rather than signal. CAR and net coincidences have been also simulated starting from the parameters calculated in sec. \ref{brightness} and sec. \ref{sec:data_anal}. They are reported as solid lines in the figure and are calculated as \cite{harada2009frequency}
\begin{subequations}
\begin{equation}
    \textrm{CAR} = \frac{\bar{p}_{si}}{\bar{p}_i \, \bar{p}_s} = \frac{\xi \bar{P}_p^2 \bar{\eta_i} \bar{\eta_s}}{\left(\xi \bar{P}_p^2 \bar{\eta_i} + d_i\right) \, \left(\xi \bar{P}_p^2 \bar{\eta_s} + d_s\right)},
\end{equation}
\begin{equation}
    N_{si}^{net} = \xi \bar{P}_p^2 \bar{\eta}_i \bar{\eta}_s \eta_{ND} R_p.
\end{equation}
\end{subequations}
Simulated and experimental values of CAR are in agreement in the whole range of pump power used. This agreement demonstrates that the main effects and phenomena involved in the generation process have been properly considered and modelled. The net coincidence rates are in agreement at low power, while at higher power the nonlinear losses have been overestimated. A perfect agreement would require a precise knowledge of all the nonlinear parameters of the material.\\
The larger CAR here measured with respect to other works \cite{rosenfeld2020mid} demonstrates that the overall system, considering both the generation and detection stages, is competitive with respect to solutions already demonstrated on the silicon platform.

\begin{table*}[h!]
\renewcommand*{\arraystretch}{1.4}
	\caption{Comparison with state of the art MIR heralded sources.}\label{T:sources}
	\begin{tabular}{|c|c|c|c|c|c|c|c|}
	\hline
\textbf{Platform} & \textbf{Process} & \textbf{Generation probability} &	\textbf{CAR}& \textbf{CAR} & $\mathbf{g^{(2)}_h(0)}$& $\eta_I$ &\textbf{Reference}\\
&  & (W$^{-2}$) & \textbf{max} & @ $N^{net}_{si} \sim$ 1 Hz & 	& ($\%$)& 	\\
\hline
 Mg:PPLN & SPDC 	& - & 180 $\pm$ 50 & - & - 	& - &\cite{prabhakar2020two}	\\
SOI & intra-modal SFWM  &	0.28 &25.7 $\pm$ 1.1	& 25.7 $\pm$ 1.1 & - &	5 &\cite{rosenfeld2020mid}	\\
SOI & inter-modal SFWM  & 0.72 $\pm$ 0.10	& 40.4 $\pm$ 0.9 & 27.9 $\pm$ 0.5 & 0.23 $\pm$ 0.08	& 59 $\pm$ 5 & This work	\\

	\hline

\end{tabular}
\end{table*}

\subsection{Heralded g$^{(2)}_h$}
\begin{figure}[h!]
\includegraphics[width=\columnwidth]{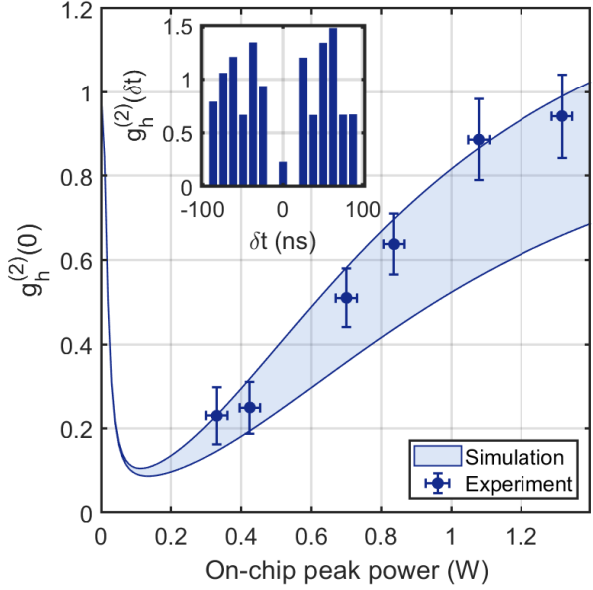}% Here is how to import EPS art
\caption{\label{fig:g2h} Comparison between the measured (blue points) and simulated (light blue area) $g_h^{(2)}(0)$ as a function of the on-chip peak power. In the inset is reported the measurement for the $g^{(2)}_h(\delta t)$ at an on-chip peak power of 0.33 W. The bins adjacent to the zero-delayed one have been removed due to the SPADs emitted photons.
}
\end{figure}

To asses the single photon nature of the emission, we measured the heralded $g^{(2)}$, that we indicate as $g^{(2)}_h$. Using the setup in Fig. \ref{fig:setup}b, we tuned the delays in order to have the signal detection on one visible SPAD coincident with the idler detection on the InGaAs SPAD. The coincidence between these two detectors, with a coincidence window $\Delta t_c =$ 2 ns, was used as the start trigger, while the detection from the remaining visibile SPAD, that we will call "delayed signal", was used as the stop trigger. In this way, we monitored the three-fold coincidences as a function of the delay $\delta t$ between the start and stop events. At the same time, we measured the two-fold coincidences between the idler and the delayed signal. We used a coincidence window of 2 ns to monitor the three-fold coincidences. The $g^{(2)}_h$ can be given as \cite{signorini2020chip}
\begin{equation} \label{g2h}
     g^{(2)}_h(\delta t) = \frac{N_{12i}(\delta t)}{N_{1i}(0) N_{2i}(\delta t)} N_i,
\end{equation}

where $1,2,i$ label respectively the first signal detector, the second signal detector (that is the delayed signal) and the idler detector. $N_{12i}$ corresponds to the three-fold coincidence counts, $N_{1i}$ and $N_{2i}$ are the two-fold coincidence counts between the idler and the signal detectors, and $N_i$ corresponds to the idler counts. We can normalize eq. \eqref{g2h} by $N_i$ and $N_{1i}(0)$, such that
\begin{equation} \label{g2hn}
    g^{(2)}_h(\delta t) = \frac{N_{12i}(\delta t)}{\langle N_{12i}(\delta t \neq 0) \rangle} \frac{\langle N_{2i} (\delta t \neq 0) \rangle}{N_{2i} (\delta t)},
\end{equation}

with $\langle N_{12i}(\delta t \neq 0) \rangle$ and $\langle N_{2i}(\delta t \neq 0) \rangle$ the average of the three-folds and two-folds coincidences for $\delta t$ different from zero. If the emission is truly at the single photon level, $g^{(2)}_h(0)$ should be lower than 0.5\cite{signorini2020chip}. The measured $g^{(2)}_h(0)$ as a function of the on-chip peak pump power is reported in Fig. \ref{fig:g2h}. For an input power of 0.33 W we measured $g^{(2)}_h(0) = 0.23(8)$, demonstrating the single photon regime of the source. The corresponding $g^{(2)}_h(\delta t)$, calculated as in eq. \eqref{g2hn}, is reported in the inset of Fig. \ref{fig:g2h}. We discarded the neighbouring bins of the zero delay bin, affected by spurious coincidences due to photon emissions from triggered silicon SPADs \cite{kurtsiefer2001breakdown}.

\begin{figure}[h!]
\includegraphics[width=\columnwidth]{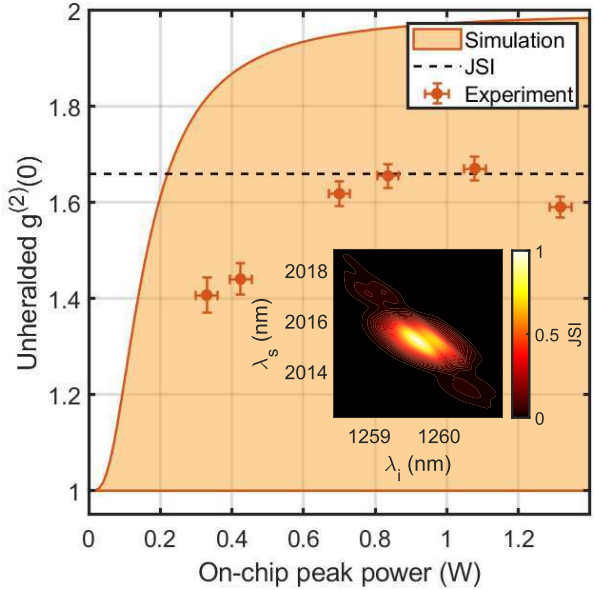}% Here is how to import EPS art
\caption{\label{fig:g2} The measured unheralded $g^{(2)}(0)$ (orange dots) is reported as a function of the on-chip peak power. We report in the inset the simulated JSI, from which we calculated the expected $g^{(2)}$ (dashed black line), that is compatible with the experiment. The measured points fall within the simulated values (light orange area), upper bounded by a source with thermal emission statistics and lower bounded by a source with poissonian emission statistics (constant $g^{(2)} = 1$). 
}
\end{figure}

To verify the goodness of the modeling introduced in sec. \ref{sec:data_anal}, we used the calculated $\xi$, $\bar{P}_p$, $\bar{\eta}_i$ and $\bar{\eta}_s$ in sec. \ref{brightness} to simulate the expected $g_h^{(2)}(0)$. Considering the general formula for the heralded second order coherence, we can write
\begin{equation}
    g_{h}^{(2)}(0) = \frac{\bar{p}_{12i} \bar{p}_i}{\bar{p}_{1i} \bar{p}_{2i}},
\end{equation}
where $\bar{p}_{12i}$ is the probability per pulse of having a three-fold coincidence. To model the experimental results, we have to consider all the possible coincidence events that may involve signal and/or noise detections. By considering all the possible events leading to a three-fold coincidence (see Supplementary Material), we can rewrite $\bar{p}_{12i}$ as
\begin{align} \label{p12i}
\bar{p}_{12i} =
& \sum_{n=2}^\infty n^2 (n-1) \wp(n) \, \bar{\eta}_1 \bar{\eta}_2 \bar{\eta}_i \eta_{ND}\\
+ & \sum_{n=1}^\infty n^2 \wp(n) \, (\bar{\eta}_1 d_2 +  d_1 \bar{\eta}_2) \bar{\eta}_i \eta_{ND}\\
+ & \frac{1}{2} \sum_{n=2}^\infty n(n-1) \wp(n) \, \bar{\eta}_1 \bar{\eta}_2 d_i \eta_{ND}\\
+ & \sum_{n=1}^\infty n \wp(n) \, \bar{\eta}_1 d_2 d_i \eta_{ND}\\
+ & \sum_{n=1}^\infty n \wp(n) \,  d_1 \bar{\eta}_2 d_i \eta_{ND}\\
+ & \sum_{n=1}^\infty n \wp(n) d_1 d_2 \bar{\eta}_i \eta_{ND}\\
+ & \, d_1 d_2 d_i \eta_{ND},
\end{align}
with $\wp(n)$ the photon number distribution. In eq. \eqref{p12i}, $\bar{\eta}_i$ is as in eq. \eqref{etajbar}, while $\bar{\eta}_1$ and $\bar{\eta}_2$ have to take into account also the effect of the beam splitter, thus, according to eq. \eqref{etajbar}, they can be written as
\begin{subequations}\label{eta12}
\begin{equation}
    \bar{\eta}_1 = \bar{\eta}_s T^2_{BS} \eta_{BS},
\end{equation}
\begin{equation}
    \bar{\eta}_2 = \bar{\eta}_s R^2_{BS} \eta_{BS},
\end{equation}
\end{subequations}
with $T_{BS}$ and $R_{BS}$ the transmission and reflection coefficients of the beam splitter, $T^2_{BS} + R^2_{BS} = 1$, and $\eta_{BS}$ modeling the losses of the beam splitter. In our case, $T^2_{BS} = R^2_{BS} = 0.5$ and $\eta_{BS} = 1$. In eqs. \eqref{eta12} we are assuming the same detection efficiency for the two visible SPADs. Considering all the events leading to a two-fold coincidence, we can rewrite $\bar{p}_{1i}$ and $\bar{p}_{2i}$ as
\begin{align}\label{pki}
    \bar{p}_{ki} = & \sum_{n=1}^\infty n^2 \wp(n) \bar{\eta}_k \bar{\eta}_i \eta_{ND} \\
    + & \sum^\infty_{n=1} n \wp(n) \left( \bar{\eta}_k d_i + d_k \bar{\eta}_i \right) \eta_{ND}\\
    + & \, d_k d_i \eta_{ND},
\end{align}
with $k = 1,2$. To be noticed that in eq. \eqref{p12i} and eq. \eqref{pki} we are neglecting events with more than one photon reaching the same detector, being unlikely with the involved transmission efficiencies (i.e. $\bar{\eta}_i$, $\bar{\eta}_1$ and $\bar{\eta}_2$ are all $\ll$1). We are also neglecting events where photon detections and dark count detections occur simultaneously on the same detector.
The photon number distribution of a squeezed source ranges between a poissonian (infinite modes emission) and a thermal (single mode emission) distribution \cite{takesue2010effects,signorini2020chip}. We solved eq. \ref{p12i} and eq. \ref{pki} for the poissonian emission,
\begin{equation} \label{ppoisson}
    \wp(n) = \frac{\mu^n}{n!} \textrm{e}^{ -\mu},
\end{equation}
and for the thermal emission,
\begin{equation}\label{pthermal}
    \wp(n) = \frac{\mu^n}{(1+\mu)^{n+1}},
\end{equation}
 where $\mu$ is the average number of pair per pulse. Eqs. \ref{ppoisson} and \ref{pthermal} define a lower and an upper boundary for $g^{(2)}_{h,sim}$. In computing $g^{(2)}_{h}$ we calculated $\mu$ as $\mu = \xi \bar{P}_p^2$ and we measured the noise affecting the three channels. $d_i$ is the same of the CAR measurements, $d_1 = 2.30 \times 10^{-6}$ and $d_2 = 2.32 \times 10^{-6}$. We simulated an area for the expected value of the $g^{(2)}_h(0)$, that is upper bounded by the thermal case and lower bounded by the poissonian case. The simulation is reported in Fig. \ref{fig:g2h}. The measured $g^{(2)}_h$ is compatible with the simulated values, confirming the reliability of the modeling. We want to stress that in this case we are not performing a fit of the measured $g^{(2)}_h$ and that the experiment and the simulation are completely independent. The experimental points in Fig. \ref{fig:g2h} are closer to the upper bound rather than to the lower one, suggesting an emission statistics closer to the thermal one. This is compatible with the unheralded $g^{(2)}$ of the signal beam \cite{signorini2020chip}, measured in Fig. \ref{fig:g2} as a function of the pump power. The unheralded $g^{(2)}$ results to be 1.67(2) at a power of 1.08 W, compatible with the simulated value of 1.66 (dashed line) calculated from the simulated joint spectral intensity (JSI) \cite{signorini2020chip,borghi2020phase}. The measured $g^{(2)}$ demonstrates that the source is closer to a thermal emission, justifying the experimental $g^{(2)}_h$. In Fig. \ref{fig:g2} we also report the simulated values for a source whose statistics is between the thermal (upper bound) and the poissonian one (lower bound). At low powers, the dark counts dominate, and in both cases the $g^{(2)}$ goes to 1. At high powers, the $g^{(2)}$ asymptotically increases to its actual value. In this way, we explain the power dependent behaviour of the experimental data. Further details about the measurement and simulation of $g^{(2)}$ are reported in Supplementary materials.

%The unheralded $g^{(2)}$ is calculated as \cite{signorini2020chip}
%\begin{equation}
 %   g^{(2)}(\delta t) = \frac{N_{12}\left( \delta t\right)}{\langle N_{12}\left( \delta t \neq 0\right)\rangle},
%\end{equation}

%where $N_{12}\left( \delta t \right)$ are the coincidence counts between the two signal detectors as a function of the delay $\delta t$ between the start and stop events.

%can be approximated with a thermal distribution \cite{husko2013multi}. Therefore, solving the series in eq. \eqref{p12i} and eq. \eqref{pki} with $\wp(n) = \mu^n / (1+\mu)^{n+1}$ leads to
%\begin{alignLetter} \label{p12i_mu}
%\bar{p}_{12i} =
%& \frac{2}{1+\mu} (3 \mu^4 + 5 \mu^3 + 2 \mu^2) \, \bar{\eta}_1 \bar{\eta}_2 \bar{\eta}_i \\
%+ & \mu (2\mu + 1) \, (\bar{\eta}_1 d_2 + d_1 \bar{\eta}_2) \bar{\eta}_i\\
%+ &  \mu^2 \, \bar{\eta}_1 \bar{\eta}_2 d_i \\
%+ &  \mu \, \bar{\eta}_1 d_2 d_i \\
%+ &  \mu \,  d_1 \bar{\eta}_2 d_i \\
%+ &  \mu d_1 d_2 \bar{\eta}_i\\
%+ & \, d_1 d_2 d_i,
%\end{alignLetter}
%and
%\begin{alignLetter}\label{pki_mu}
 %   \bar{p}_{ki} = & \mu (2 \mu + 1) \bar{\eta}_k \bar{\eta}_i \\
%    + & \mu \left( \bar{\eta}_k d_i + d_k \bar{\eta}_i \right)\\
%    + & \, d_k d_i,
%\end{alignLetter}.

\section{Conclusions}\label{sec:conclusions}
In this work, we demonstrated a heralded single photon source beyond 2 $\mu$m based on inter-modal SFWM on a silicon chip. This source has two main peculiarities: the discrete band generation and the large detuning between the signal and idler photons. The discrete band generation removes the need for tight filtering to select idler and signal wavelengths, and the generated photons experience a higher transmission with respect to standard continuous band sources, witnessed by the high experimental $\eta_I = 59(5) \, \%$. The large detuning has two advantages: on one side, it enables an easier pump and nonlinear noise rejection; on the other side, it allows to generate the herald photon in the NIR, benefiting of an efficient detection technology. As a last advantage, this heralded source based on inter-modal SFWM requires a common C-band pump laser, easier to be integrated and operated on a silicon chip.
We performed a complete characterization of the source. We demonstrated the sub-poissonian statistics of the source by measuring $g^{(2)}_h(0) = 0.23(8)$. We characterized the CAR, finding a maximum value of 40.4(9), and the generation probability per pulse, with a measured value of 0.72(10) W$^{-2}$. These performances are competitive with other reported silicon sources of MIR photons (Table \ref{T:sources}) demonstrating the promising perspectives of inter-modal SFWM for bright and efficient sources of correlated photons beyond 2 $\upmu$m. The source can be significantly improved by reducing the propagation losses and optimizing the matching between the signal and upconverter bandwidths. With this work we demonstrate a new approach to MIR quantum photonics, providing a high quality source of quantum light beyond 2 $\mu$m without the need of MIR technologies. This result paves the way towards low cost, efficient and integrated solutions for quantum photonics beyond 2 $\upmu$m, offering new opportunities to the developing field of MIR photonics.

% NB: NESSUNO HA MAI RIPORTATO COINCIDENZE A TRE CHE COINVOLGESSERO IL MIR. 
% NB: stressare di più heralding efficiency
% NB: stressare di più che con inter-modale si possono separare idler e signal più facilmente su chip
% NB: i dark counts si dimezzano una volta inserito il beam splitter
% NB: noise due to Raman in Fiber

% Figures should be put into the text as floats. 
% Use the graphics or graphicx packages (distributed with LaTeX2e).
% See the LaTeX Graphics Companion by Michel Goosens, Sebastian Rahtz, and Frank Mittelbach for examples. 
%

% Tables may be be put in the text as floats.
% Here is an example of the general form of a table:
% Fill in the caption in the braces of the \caption{} command. Put the label
% that you will use with \ref{} command in the braces of the \label{} command.
% Insert the column specifiers (l, r, c, d, etc.) in the empty braces of the
% \begin{tabular}{} command.
%
% \begin{table}
% \caption{\label{} }
% \begin{tabular}{}
% \end{tabular}
% \end{table}

\section*{SUPPLEMENTARY MATERIAL}
See supplementary material for further details about the experimental setup, the measurements and the theoretical calculations.

\begin{acknowledgments}
This work was partially supported by grants from Q@TN provided by the Provincia Autonoma di Trento. The authors acknowledge HC Photonics, which fabricated the PPLN crystals used for the upconversion system. S.S. wants to thank Dr. Massimo Borghi, for fruitful discussions and precious suggestions, and Mr. Davide Rizzotti for its careful revision of the manuscript.
\end{acknowledgments}

\section*{DATA AVAILABILITY}
The data that support the findings of this study are available from the corresponding author
upon reasonable request.
\section*{REFERENCES}
\bibliography{TESTO}

\end{document}